\documentclass[journal]{IEEEtran}
\usepackage{amsmath,amsfonts}
\usepackage{algorithmic}
\usepackage{algorithm}
\usepackage{array}
\usepackage[caption=false,font=normalsize,labelfont=sf,textfont=sf]{subfig}
\usepackage{textcomp}
\usepackage{stfloats}
\usepackage{url}
\usepackage{verbatim}
\usepackage{graphicx}
\usepackage{cite}

\usepackage{amssymb}
\usepackage{empheq}

\begin{document}

\title{Hall Effect, Magnetoresistance, and Current Distribution in Quench Heaters}

\author{J.~Rysti
\thanks{Manuscript received July 27, 2024; revised xx xx, 2024.}
\thanks{The author was with CERN, 1211 Geneva 23, Switzerland. . He is now with Department of Applied Physics, Aalto University, POB 15100, FI-00076, AALTO, Finland (e-mail: juho.rysti@aalto.fi). }
}



\maketitle

\begin{abstract}
Quench heaters are often an essential part of protecting a superconducting accelerator magnet during a quench. Their purpose is to spread the quench throughout the coil as quickly as possible. They are located in areas of high magnetic fields and are thus prone to magnetoresistive phenomena and the Hall effect. Such influences can cause currents to distribute unevenly in the heaters, which results in uneven heating. This can reduce the effectiveness of the heaters and even endanger them due to excessive local heating. Also, the heater geometry itself can be the cause of uneven current density.

In this paper we investigate by numerical simulations the importance of the magnetic effects on quench heater performance and whether they should be taken into account in the design. The main interest is in the Hall effect, which was perceived as the most likely source of trouble for the design of quench heaters. We use a simple phenomenological approach for modeling the Hall effect, utilizing values from the literature for the Hall coefficients. Magnetoresistance is also considered and the impact of heater geometry on current distributions is briefly visited.

The conclusion of this research is that magentoresistance plays an insignificant role in the functioning of quench heaters. The Hall effect can clearly be more influential, but nevertheless should not pose any problems in most cases. Current distributions due to heater geometry should be take into consideration in the design phase and, if needed, take measures to equalize the current density by using, for example, copper cladding in appropriate locations.
\end{abstract}


\section{Introduction}
\IEEEPARstart{D}{uring} the quench of a superconducting magnet, it may be necessary to spread the quench throughout the conductor as quickly as possible to avoid high hot spot temperature in the initial quench location. Quench heaters are often used for this purpose \cite{LHCProtection}. They can be manufactured in a variety of shapes and configurations, but perhaps the simplest kind of quench heater is a straight thin strip of stainless steel glued on top of the electrical insulation of the coil. After detection of a quench, current is quickly passed through the quench heaters from a capacitor bank and their temperature is increased rapidly by Joule heating. The heat diffuses into the cable and raises its temperature above the current sharing temperature of the superconductor, thus causing it to quench.

However, the power density of quench heaters is limited by the maximum allowed voltage across the strip, which is typically around 1~kV. This is a problem especially for long magnets, where long heater strips with large resistance are needed. The situation can be remedied by having heating stations at regular intervals instead of continuous heaters along the entire length of the magnet. Protection of the magnet then partially relies on the propagation of quench between the stations. This type of system can be produced, for example, by copper plating the parts of the steel strips, where heating is not wanted. Current prefers to travel in the copper, which at low temperatures has a conductivity roughly 1000 times higher than stainless steel. Another possibility is to have a wide, and possibly thick, steel strip and construct the heating stations by making the strip narrow and/or thin at the desired heating station locations. Heating power density is then locally increased by the smaller cross-sectional area. In this type of design heating stations can be, for example, S-shaped or consist of multiple U-shaped turns so the current is forced to provide more uniform heating across the conductors. The problem with these types of heaters is that the power density is often unequal, as will be demonstrated later, and the power of the heating station is limited, since the entire heater strip is made of the same resistive material.

In some cases it may also be beneficial to minimize the total surface area of the heater system. This has turned out to be desirable in cases, where the heaters do not have support against the coil surface, such as on the inner layer of coils. It seems that superfluid helium is able to enter underneath the heater strips. During a quench the helium cannot exit through the Kapton insulation layer quickly enough and the ensuing pressure tends to detach the heater trace from the coil \cite{MQXF}. For this purpose copper plating offers possibilities, since the plated parts can be narrower and the total surface area is reduced. But, as we shall see, the connection between the narrower copper-plated sections and the wider steel parts must be done carefully to guarantee good heater performance.

Quench heaters experience high magnetic fields and large field variations. In order to ensure the efficient functioning of the heaters, the magnitude of effects due to magnetic fields on the current distribution should be known. If found to be excessively strong, they can be alleviated by heater design features. In this paper we investigate the importance of both the Hall effect and magnetoresistance on the distribution of current in quench heaters. The main interest is in the Hall effect, which was expected to be of more importance and because no previous studies on the subject were found.

\begin{figure}[!t]
\centering
\includegraphics[width=\columnwidth]{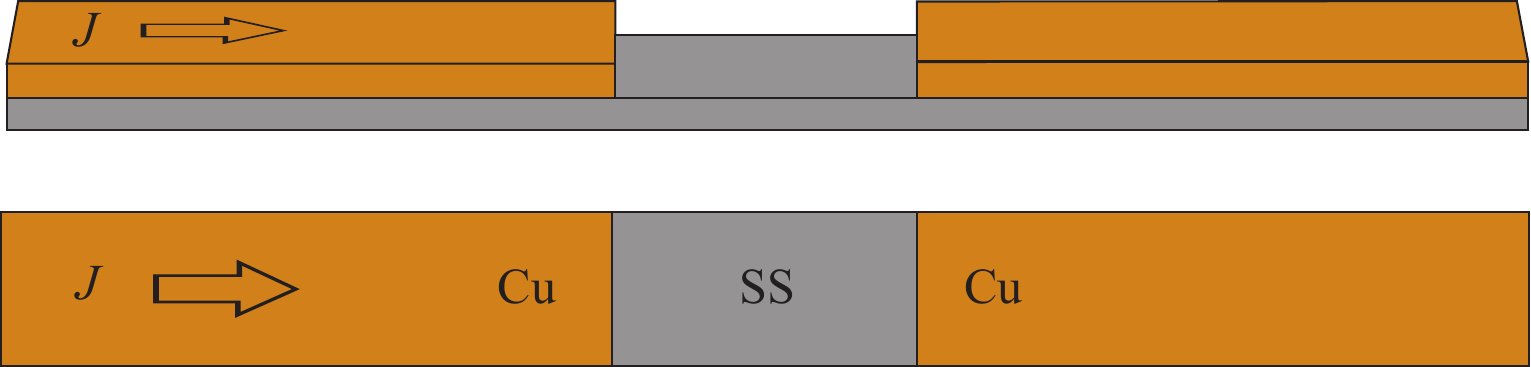}
\caption{Section of a copper plated quench heater strip geometry showing one heating station in the middle. The full 3D system is shown in the upper figure and the 2D approximation in the lower, where the materials extend to infinity in the direction perpendicular to the figure. The relative dimensions in the 3D geometry have been exaggerated. Current flows along the strip.}
\label{Fig_QHGeometry}
\end{figure}

In this article we mainly consider quench heaters with heating stations made by copper-plating stainless steel strips. The geometry of such quench heaters is illustrated in Fig.~\ref{Fig_QHGeometry}.
The 2D approximation (lower figure) corresponds to the situation, where the copper and stainless steel parts extend to infinity in the direction perpendicular to the figure. In reality steel covers the entire length of the heater strip and copper is plated on top of the steel in areas, where heating is not wanted (upper figure). The thicknesses compared to the other dimensions have been exaggerated. In reality the thickness of stainless steel and copper are typically 25~\textmu m and 5~\textmu m, respectively. The width of the heater strips are usually between 10-40~mm and the length can be up to several meters. The 2D approximation is assumed to be good, since the conductivity of copper at low temperatures dominates that of stainless steel by a factor of about 1000.

\section{Computational methods}
\label{Sec_ComputationalMethods}

In the simulations we assumed the system to be stationary. Whether this is a good approximation for modeling current distributions in quench heaters is basically determined by the time scale of current changes of the system compared to the charge relaxation time $\tau = \epsilon/\sigma$, which is the ratio of the permittivity and conductivity of the material. Using the vacuum permittivity, copper at low temperature has $\tau \sim 10^{-21}$~s and steel has $\tau \sim 10^{-18}$~s. Typical time scales of quench heater operations are around ten milliseconds. It is thus a reasonable assumption that the relative current distribution remains roughly constant during the heater powering. Temperature rise in the heaters will, of course, change the conductivity, but this can be modeled with the same stationary equations, since the time scales of the resistivity changes are much longer than the charge relaxation time. However, since the change in the resistivity of steel as a function of temperature is moderate anyway (RRR $\approx 1.5$) and we are only interested in the magnitude of the current distribution effects, we have neglected temperature variations in the present study and only examine the truly static cases.

The equations describing stationary electric currents are
\begin{empheq}[]{align} 
&\nabla \cdot \vec{J} = 0 \label{Eq_Jconserv} \\
&\vec{J} = \sigma \vec{E} \label{Eq_OhmLaw} \\
&\vec{E} = -\nabla V \label{Eq_ElecPot},
\end{empheq}
where $\vec{J}$ is the current density, $\sigma$ is the conductivity (reciprocal of resistivity $\rho$), $\vec{E}$ is the electric field, and $V$ the electric potential. The first equation is the conservation of current, the second one is a generalization of Ohm's law, and the last one is the definition of the electric potential. In addition to these, we need to specify the appropriate boundary conditions. On most surfaces we assume electrical insulation, which means the current density normal to the surface is zero ($\hat{n} \cdot \vec{J} = 0$). We define appropriate voltage boundary conditions, where we want the current (current density) to enter and exit the system. Modeling the Hall effect requires some additional attention, which is discussed in more detail in Sec.~\ref{Sec_HallEffect}.

The equations were solved with the finite element method, which is useful for complicated geometries. We have used COMSOL Multiphysics and Matlab for this purpose. For the current distribution and magnetoresistance simulations we only present the results of 2D computations. They were verified by comparing to a 3D model, where a copper layer was placed on top of the steel strip (Fig.~\ref{Fig_QHGeometry}, upper). For the Hall effect we discuss the 3D model, but also show only the results of 2D simulations, as the results are practically identical for the two cases.

\section{Heater geometry}

\begin{figure}[!t]
\centering
\includegraphics[width=\columnwidth]{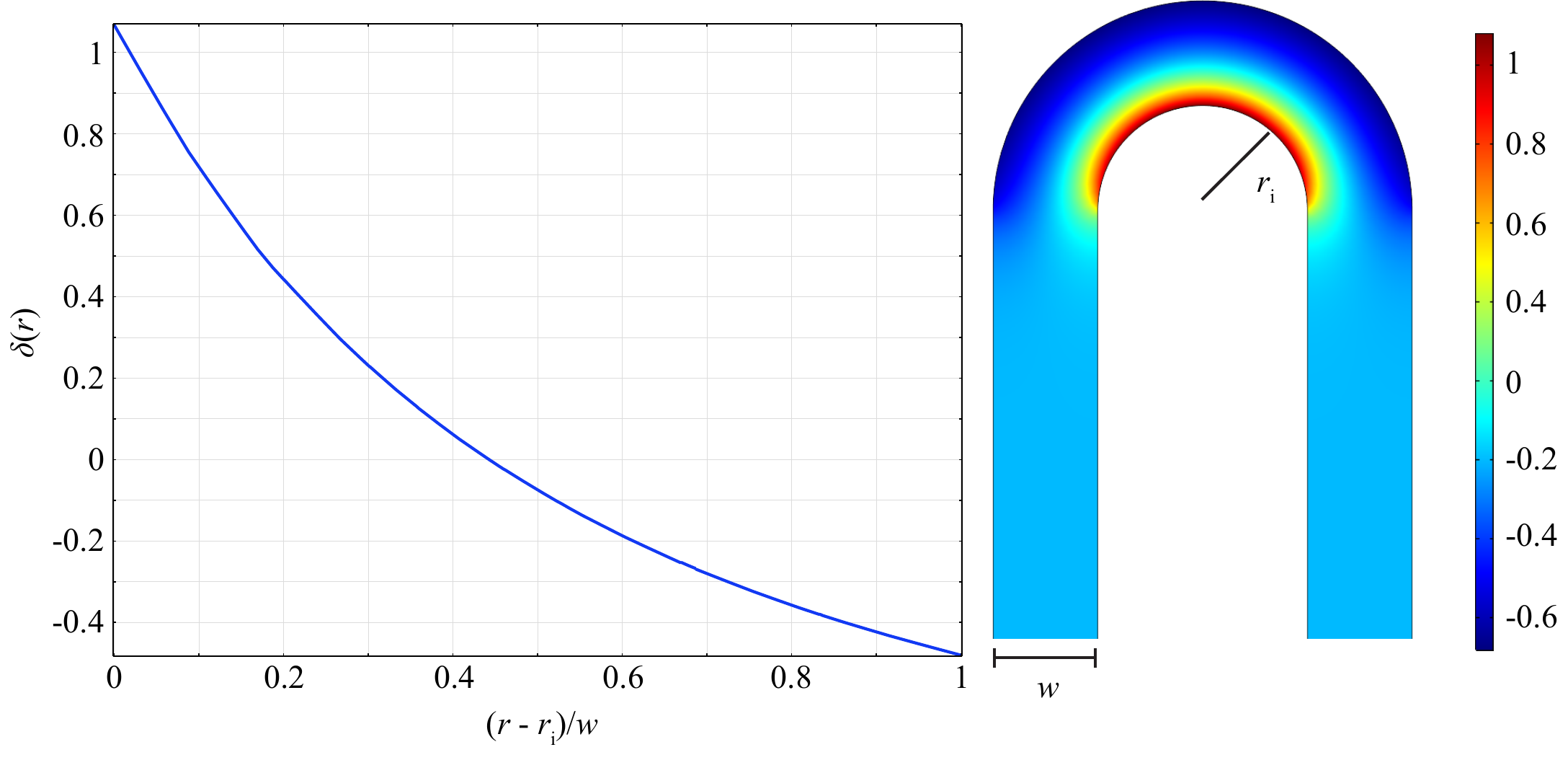}
\caption{Relative variation in Joule heating in a quench heater consisting of a uniform straight strip forming a U-shape. The colors depict the value of $\delta(x,y)$, defined in Eq.~(\ref{Eq_RelativeJSquare}). The ratio of the inner radius and width is $r_\textrm{i}/w=1$. The inset shows $\delta$ on a vertical cut line at the middle of the turn.}
\label{Fig_UTurn}
\end{figure}

First we show how the geometry of quench heaters can affect the current distribution, and thus the heating. Of course if the heater design is very simple with just a straight strip with a constant width, current is uniformly distributed (when neglecting magnetoresistance). However, if the constant-width strip makes a U-turn, current will distribute unevenly in the turn. Fig.~\ref{Fig_UTurn} shows the results for a heater strip of width $w$, which makes a U-turn with an inner radius $r_\textrm{i}=w$.
The color scale depicts the relative difference of current density squared $J^2$ compared to the square of the average current density $\langle J \rangle ^2$:
\begin{equation} \label{Eq_RelativeJSquare}
\delta (x,y) \equiv \frac{J(x,y)^2-\langle J \rangle ^2}{\langle J \rangle ^2}.
\end{equation}
Joule heating is proportional to $J^2$, so $\delta$ is representative of the variation in the heating power density. The result in Fig.~\ref{Fig_UTurn} does not depend on the absolute size of the strip, but only on the relative dimensions. In the case shown, $J^2$ is 108\% larger along the inner edge of the turn than the average current density, and 48\% smaller on the outer edge. When changing the relative dimensions, the maximum $\delta$ is almost a linear function of the ratio $w/r_\textrm{i}$. A power-law fit between $w/r_\textrm{i} = 0$ and $w/r_\textrm{i} = 4$ gives $\delta_\textrm{max} = 1.090 (w/r_\textrm{i})^{1.073}$. For the minimum a hyperbolic fit was found to give good correspondence: $\delta_\textrm{min} = -0.906 (w/r_\textrm{i})/(0.8911+(w/r_\textrm{i}))$. For heater strips with turns, one should make sure the additional heating on the inner edge does not results in overheating.

In the above case we have assumed a constant conductivity and neglected any changes due to heating. Temperature rise of the heater tends to equalize the current density and thus alleviates the situation, since it increases the resistivity. But the resistivity of steel changes only by a factor of about 1.5 between 2~K and 300~K. In the case of copper heaters, where resistivity changes by a factor of about 100, this may be of more significance. Such heaters could be made, for example, of copper strips with heating stations formed by narrow sections of alternating U-turns.

\begin{figure}[!t]
\centering
\includegraphics[width=\columnwidth]{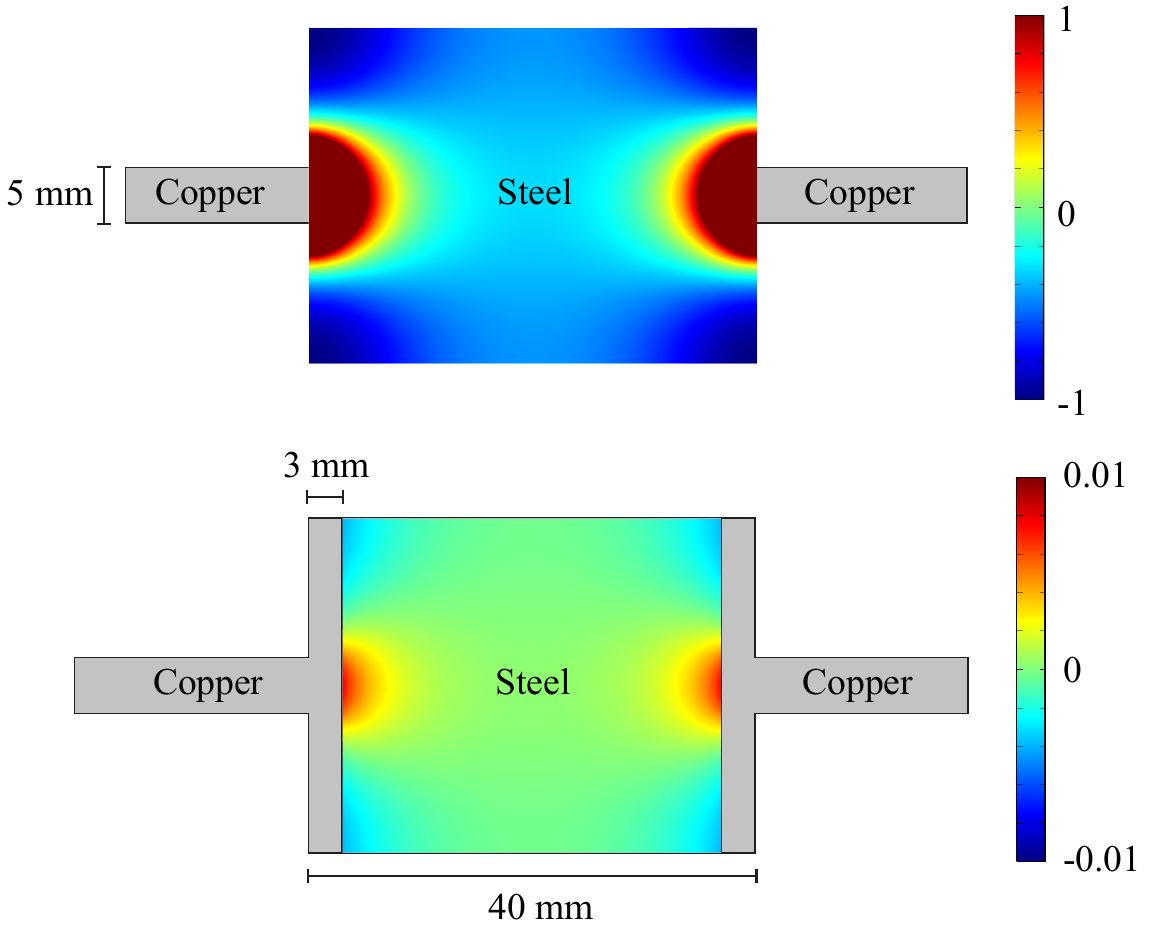}
\caption{Relative variation in Joule heating, $\delta$, in quench heaters with stainless steel heating stations connected by narrow copper plated leads. In the upper case the leads are directly connected to the heating station, while in the lower case the copper plating extends 3~mm to cover the height of the heating station (30 mm). The color range in the upper figure is limited to $\pm 100$\%. In the lower figure the colors cover the entire range of variation.}
\label{Fig_ExtendedCopper}
\end{figure}

Next we consider a slightly more complicated geometry with stainless steel heating stations of dimensions $40 \times 30$~mm connected by 5~mm wide copper plated leads. This heater geometry and the dimensions were motivated by a desire to minimize the total surface area of the heater strips for the inner layer heaters of the MQXF magnet \cite{MQXFProtection}. We compare two cases; one where the copper plated leads are attached directly to the heating stations and a case, where the copper plating extends the entire height of the heating station for a length of 3~mm. Fig.~\ref{Fig_ExtendedCopper} shows the two geometries and the results for the variation in heating.
Again, the color coding represents the relative current density squared, defined in Eq.~(\ref{Eq_RelativeJSquare}). One should note that in the upper figure the color scale is limited to $\pm~100\%$ and includes areas, where the current density varies even more (maximum $\delta>100$). If the copper plating is extended to cover also a portion of the width of the heating station, as in the lower figure, currents are much more uniformly distributed. The variation in heating is less than $\pm 1$\%. The high conductivity of copper compared to steel allows the currents to spread easily along the entire width of the heaters. It is clear that the copper extension is crucial in cases similar to that of Fig.~\ref{Fig_ExtendedCopper} to ensure even heating.

One may devise other possible ways to equalize the current distribution in heaters, such as connecting the copper plated leads to the diagonally opposite corners of the heating station, but this arrangement in fact turns out to be worse than connecting the leads to the midpoints on both sides. This is because the current has less directions to move to when entering the steel from copper.

\section{Magnetoresistance}
\label{Sec_Magnetoresistance}

Magnetoresistance is a general term for effects, which change the resistivity of a material as a function of magnetic field. In this section we focus on effects other than the Hall effect, which will be considered in the next section. To model magnetoresistance, we simply make the conductivity in Eq.~(\ref{Eq_OhmLaw}) depend on the magnetic field $\sigma = \sigma(B)$ and assign some appropriate field profile on the heater. The field dependence of $\sigma$ is obtained from the literature. We do not differentiate here between the transverse ($\vec{B}\perp\vec{J}$) and longitudinal ($\vec{B}\parallel\vec{J}$) magnetoresistance, but consider the magnetoresistivity to be isotropic. Variations in current distribution are produced by differences in the local magnetic fields in the heaters. Again, temperature variations appear in the heaters, but this only tends to even the variations in the current density. Therefore we have neglected such effects and instead show the worst-case scenario, where temperature is uniform. The conductivity of stainless steel is almost independent of magnetic field, since it is dominated by the impurities in the metal system. The properties of copper, on the other hand, exhibit clear dependence on magnetic field. For the field dependence of $\rho_\textrm{Cu}$ we have used the fits provided by NIST \cite{Drexler}.

\begin{figure}[!t]
\centering
\includegraphics[width=\columnwidth]{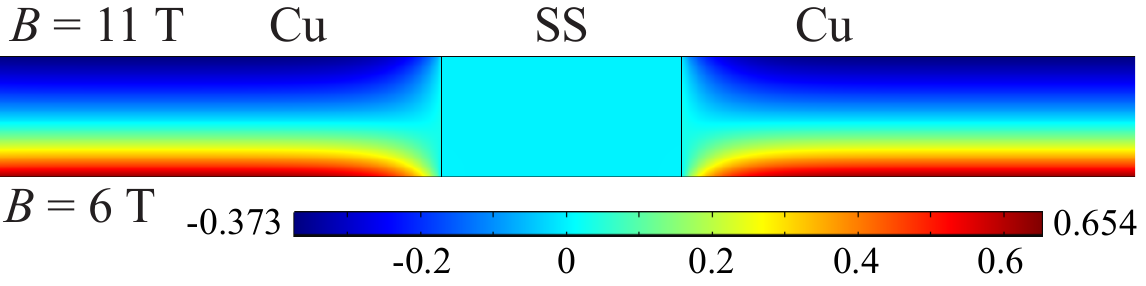}
\caption{Variation in Joule heating, $\delta$, in a straight 20~mm wide quench heater strip when the magnetic field varies linearly in the vertical direction between 6~T and 11~T. The heating station is 40~mm $\times$ 20~mm.}
\label{Fig_Magnetoresistance}
\end{figure}

\begin{figure}[!t]
\centering
\includegraphics[width=\columnwidth]{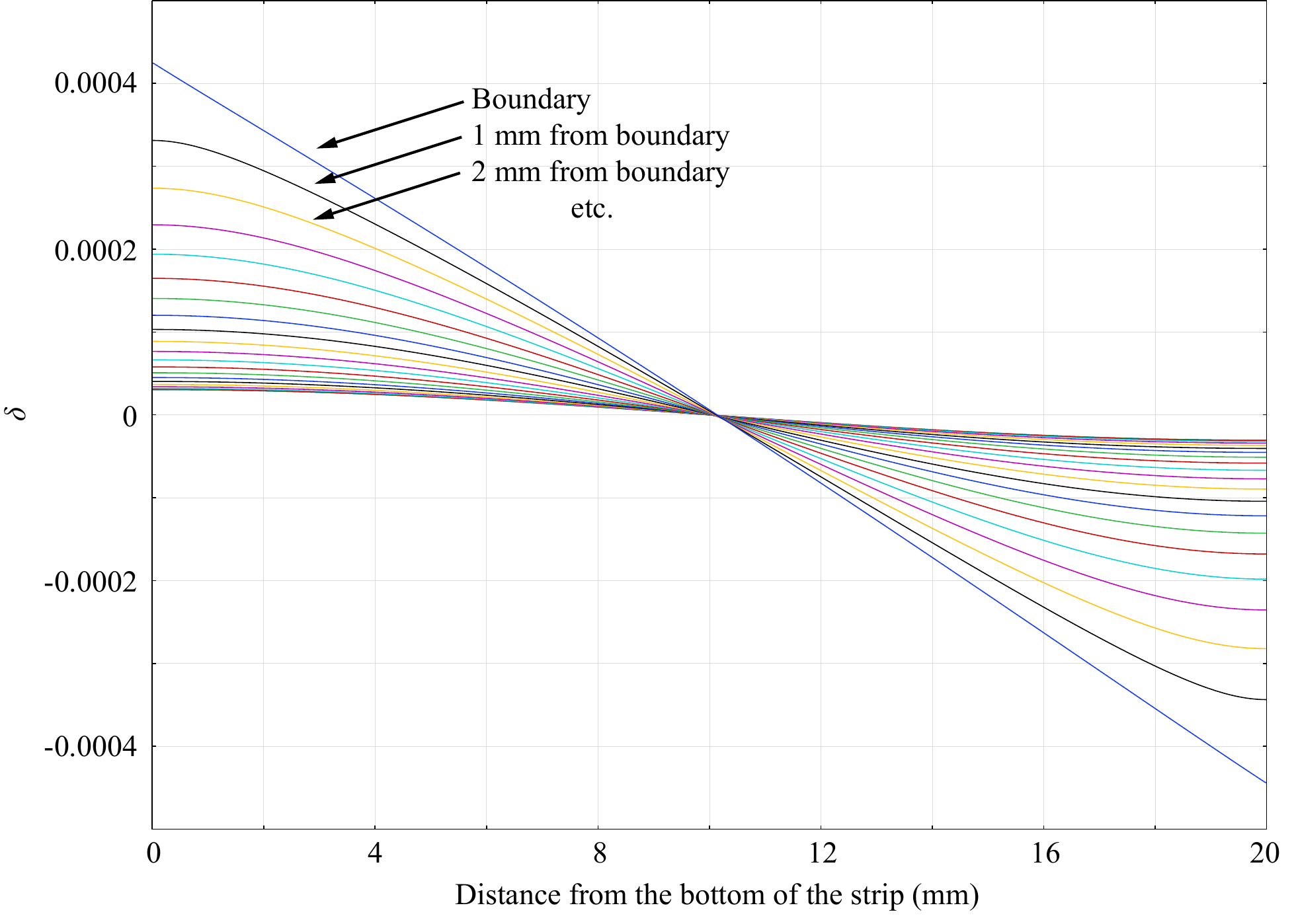}
\caption{Relative current density squared, $\delta$, in stainless steel at vertical slices starting from the boundary every 1~mm to the midpoint of the heater. The magnetic field varies linearly in the vertical direction between 6~T and 11~T. The geometry of the heater is shown in Fig.~\ref{Fig_Magnetoresistance}.}
\label{Fig_MagnetoresistancePlotSS}
\end{figure}

\begin{figure}[!t]
\centering
\includegraphics[width=\columnwidth]{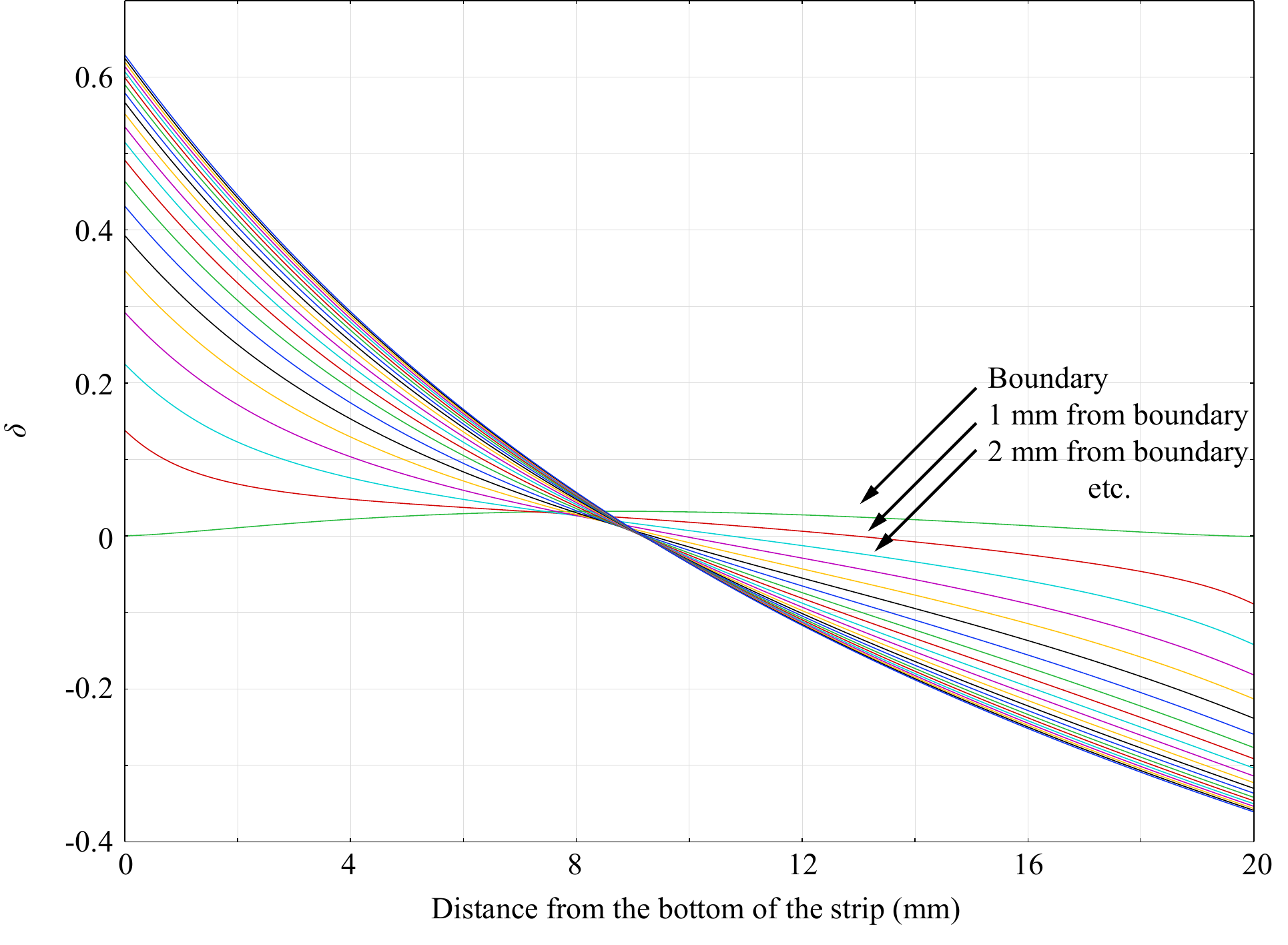}
\caption{Relative current density squared, $\delta$, in copper at vertical slices starting from the boundary every 1~mm to 20~mm away from the heating station. The magnetic field varies linearly in the vertical direction between 6~T and 11~T. The geometry of the heater is shown in Fig.~\ref{Fig_Magnetoresistance}.}
\label{Fig_MagnetoresistancePlotCu}
\end{figure}

As a model example we take the magnetic field of the MQXF quadrupole magnet of the High Luminosity LHC upgrade \cite{MQXFProtection}. We only consider the case of a straight heater strip. Between the upper and lower edges of 20~mm wide heaters the magnetic field of the MQXF varies by 5~T. We assume this variation to be linear, but one should note that $\sigma_\textrm{Cu}(B)$ is not linear. The actual values of conductivities used are $\sigma_\textrm{Cu,min}=1.8\cdot 10^9$~S/m, $\sigma_\textrm{Cu,max}=2.9\cdot 10^9$~S/m, and $\sigma_\textrm{SS}=2.0\cdot 10^6$~S/m \cite{Drexler,advanced1977handbook}. This results in a conductivity difference of about 60\% in copper. Since current density is proportional to the conductivity, also $J$ varies by the same amount between the edges of the heater. The result of the simulation is shown in Fig.~\ref{Fig_Magnetoresistance}, where again the colors denote the values of $\delta$, defined in Eq.~(\ref{Eq_RelativeJSquare}).
Even though there is appreciable variation in current density in the copper leads, it quickly re-distributes itself evenly as it approaches the stainless steel heating station. This is due to the much smaller conductivity of steel. Fig.~\ref{Fig_MagnetoresistancePlotSS} gives the current density in the stainless steel at various distances from the copper leads.
We see the maximum variation in $\delta$ is only 0.04\%, occurring at the boundary of the two materials. Fig.~\ref{Fig_MagnetoresistancePlotCu} gives the same plots in copper.
We have used a rather large RRR of 150, which gives about 60\% difference in conductivity between the lower edge ($B=6$~T) and the upper edge ($B=11$~T). With RRR$=50$ the difference would be 40\%.

At the same time as we are studying the non-uniformity due to magnetic fields, we also see what a variation in the copper RRR would cause to the current density distribution in the steel. The effect is essentially the same; at low temperatures the conductivity is no longer uniform. In the case studied the conductivity changed by about 60\% between the edges of the strip, so we would obtain the same results if we changed the RRR by the same amount at constant magnetic field. We can thus note that the performance of quench heaters is not very sensitive to the uniformity of copper RRR.

\section{Hall effect}
\label{Sec_HallEffect}

Hall effect is the production of voltage across an electrical conductor in a magnetic field when current is passed through the conductor. The so-called Hall voltage (Hall field) is perpendicular to both the magnetic field and the current (current density). Classically this can be understood to be a consequence of the Lorentz force acting on the charged current carriers, which creates a charge imbalance on the two sides of the conductor. Our model of the Hall effect is discussed here in some detail and the derivations of the relevant equations are described, since this includes some subtleties and no directly applicable references were found in the literature.

In this study we use a simple approach, where we consider the conductivity (or resistivity) of the material anisotropic. That is, the conductivity becomes a tensor quantity, whose elements depend on the magnetic field and the so-called Hall coefficient. Thus Eq.~(\ref{Eq_OhmLaw}) becomes $\vec{J} = \bar{\sigma} \vec{E}$, where the bar above the conductivity denotes a tensor quantity.

Modeling the Hall effect in this manner is valid when neglecting time transient phenomena of the current carriers, so their motion is assumed to be stationary. The general form of the equations and the conductivity tensor can be derived from the simple Drude model of electrical conduction, when assuming a steady state for the current carriers. This approximation should be good for the slowly varying currents of quench heaters. The relaxation time of electrons in the Drude model is $\tau = \sigma_0 m/ne^2$, where $m$ is the electron mass, $n$ is the number density of conduction electrons, and $e$ is the electron charge. For copper at low temperature this is around $10^{-12}$~s and for steel it is $\sim 10^{-15}$~s. Also, recall the very short charge relaxation times of copper and steel, mentioned in Sec.~\ref{Sec_ComputationalMethods}.

\subsection{Conductivity matrix}

Since the Hall field $E_\textrm{H}$ "balances the Lorentz force", it is expected to be proportional both to the applied field $B$ and the current density $J$. The Hall coefficient $R_\textrm{H}$ is then defined as
\begin{equation} \label{Eq_HallCoeffDef}
R_\textrm{H} = \frac{E_\textrm{H,y}}{J_\textrm{x} B_\textrm{z}},
\end{equation}
where the subscripts x, y, and z signify the vector components of the given quantities in the cartesian coordinate system. More generally, in non-linear cases, the Hall coefficient is defined with the derivative of the Hall field with respect to the magnetic field.

From a classical point of view the Hall coefficient is always expected to be negative, since the current carriers are negatively charged electrons. In reality the situation is not so simple. From the quantum mechanical perspective, current can equally well be transported by positively charged holes in the valence electron system. Indeed, the Hall coefficient has been observed to be positive for many materials. Deriving the Hall coefficients from first principles is extremely difficult. We take the measured values of $R_\textrm{H}$ from the literature and thus resort to a phenomenological approach for modeling the electrical conduction.

Next we derive the conductivity tensor for the Hall effect, needed in Eq.~(\ref{Eq_OhmLaw}). The total electric field $\vec{E}$ is now composed of an ohmic component $\vec{E}_\Omega$ and a Hall component $\vec{E}_\textrm{H}$: $\vec{E} = \vec{E}_\Omega + \vec{E}_\textrm{H}$. The ohmic component is simply $\vec{E}_\Omega = \bar{\rho_0} \vec{J}$, where $\bar{\rho_0}$ is a diagonal matrix with the zero-field scalar resistivity $\rho_0$ as its components. This means we model the non-Hall part of the conductivity as isotropic. In principle $\rho_0$ could be replaced by the measured field-dependent resistivities, which were used in Sec.~\ref{Sec_Magnetoresistance}, but in studying the Hall effect this is not important. The Hall component of the electric field is $\vec{E}_\textrm{H} = R_\textrm{H} ( \vec{J} \times \vec{B} )$. This is just the general statement that the Hall field is perpendicular to both $\vec{J}$ and $\vec{B}$ and its magnitude is proportional to the Hall coefficient according to Eq.~(\ref{Eq_HallCoeffDef}).

Writing the cross product as a matrix and vector multiplication using the cartesian components of $\vec{J}$ and $\vec{B}$, and adding $\vec{E}_\Omega + \vec{E}_\textrm{H}$, we find
\begin{equation}
 \begin{pmatrix}
  E_\textrm{x}  \\
  E_\textrm{y} \\
  E_\textrm{z}
 \end{pmatrix}
  = \underbrace{
 \begin{pmatrix}
  \phantom{-}\rho_0 & \phantom{-}R_\textrm{H}B_\textrm{z} & -R_\textrm{H}B_\textrm{y} \\
  -R_\textrm{H}B_\textrm{z} & \phantom{-}\rho_0 & \phantom{-}R_\textrm{H}B_\textrm{x} \\
  \phantom{-}R_\textrm{H}B_\textrm{y} & -R_\textrm{H}B_\textrm{x} & \phantom{-}\rho_0
 \end{pmatrix}}_{\equiv \bar{\rho}}
 \begin{pmatrix}
  J_\textrm{x}  \\
  J_\textrm{y} \\
  J_\textrm{z}
 \end{pmatrix}
\end{equation}
By comparing this to $\vec{E} = \bar{\rho} \vec{J}$, we can identify the resistivity matrix (tensor) $\bar{\rho}$. For simplicity, from here on we consider a uniform magnetic field, which is directed along the $z$-axis (we set $B_\mathrm{x}=B_\mathrm{y}=0$). By taking the inverse of the resistivity matrix in this case, we obtain the following conductivity matrix:
\begin{equation}
 \bar{\sigma}= \frac{\sigma_0}{1+h^2}
 \begin{pmatrix}
  \phantom{-}1 & -h & 0\phantom{^2} \\
  \phantom{-}h & \phantom{-}1 & 0\phantom{^2} \\
  \phantom{-}0 & \phantom{-}0 & 1 + h^2
 \end{pmatrix},
\end{equation}
where $h \equiv \sigma_0 R_\textrm{H} B_\textrm{z}$ and $\sigma_0 \equiv 1/\rho_0$. With this and Eqs.~(\ref{Eq_Jconserv})--(\ref{Eq_ElecPot}) we can simulate the Hall effect. In 2D the conductivity tensor is further reduced to a $2 \times 2$ matrix.

\subsection{Thin film effects}

Since quench heaters are made of thin metal strips, one might worry about the possible thin-layer effects on the Hall coefficients. In this subsection we argue that the thin-layer effects can be ignored in our simulations, where we are only interested in the magnitude of the Hall effect. MacDonald and Sarginson showed that if the magnetic field is in the direction along the film, the Hall coefficient is reduced \cite{MacDonald}. Therefore the bulk value of the Hall coefficient is the worse-case scenario compared to the thin-layer value. If the field is normal to the film, however, the Hall effect can be enhanced. Sondheimer considered this case and showed that the thin film Hall coefficient is about the same as the bulk value if either of the dimensionless parameters $\beta$ or $k$ satisfy $\beta \gtrsim 1$ or $k \gtrsim 0.5$, where $\beta \equiv d/r$ and $k \equiv d/\lambda$ \cite{PhysRev.80.401}. Here $d$ is the thickness of the film, $r$ is the radius of the circular orbit of an electron in the magnetic field (cyclotron radius), and $\lambda$ is the electron mean free path. The radius and the mean free path can be written in terms of the Fermi momentum $p_\textrm{F}$ of the conduction electrons and the bulk conductivity $\sigma_0$ of the system: $r = p_\textrm{F}/e B$ and $\lambda = 3 \pi^2 \hbar^3 \sigma_0/e^2 p_\textrm{F}^2$. The Fermi momentum, on the other hand, is obtained from the number density $n$ of conduction electrons, $p_\textrm{F} = (3 \pi^2 \hbar^3 n)^{1/3}$.

Let us estimate the factors $\beta$ and $k$ for copper and stainless steel. The number density of conduction electrons in metals is approximately equal to the atomic number density. Thus, for copper and iron $n \approx 8 \cdot 10^{28}/\textrm{m}^3$. The typical thickness of stainless steel quench heater strips is 25~\textmu m. The copper plating on the steel can be as thin as 5~\textmu m. Using a value of $\rho_\textrm{SS} = 5 \cdot 10^{-7}~\Omega \cdot \textrm{m}$ for the AISI 316 stainless steel resistivity \cite{advanced1977handbook}, the two parameters are $\beta\approx 30$ and $k\approx 2\cdot 10^{4}$ at $B=10~\textrm{T}$ and $\beta\approx 3$ and $k\approx 2\cdot 10^{4}$ at $B=1~\textrm{T}$. It is clear that the stainless steel Hall coefficient is very close to the bulk value. For copper this is not as clear due to the smaller thickness and higher conductivity. Assuming a high residual-resistivity ratio of 150, we take the 1.9~K copper resistivities at 10~T, 1~T, and 0.1~T as $\rho_\textrm{Cu} = 5 \cdot 10^{-10}~\Omega \cdot \textrm{m}$, $\rho_\textrm{Cu} = 1.4 \cdot 10^{-10}~\Omega \cdot \textrm{m}$, and $\rho_\textrm{Cu} = 1 \cdot 10^{-10}~\Omega \cdot \textrm{m}$, respectively \cite{Drexler}. The corresponding values of the dimensionless parameters for the three cases are $\beta\approx 6$ and $k\approx 4$, $\beta\approx 0.6$ and $k\approx 1$, and $\beta\approx 0.06$ and $k\approx 0.8$. We see that particularly at low fields the bulk approximation approaches the limits of its validity. But at the lowest fields the Hall effect is small anyway (proportional to $B$). Furthermore, even when $\beta < 1$ and $k < 0.5$ the increase of $R_\textrm{H}$ is not extremely large. Only for $\beta < 1$ and $k \rightarrow 0$ does the value of the Hall coefficient increase rapidly. When $k=0.1$, $R_\textrm{H}$ is 1.5 times that of the bulk value, for $k=0.02$ it is 3 times the bulk. We thus conclude that for our purposes thin-layer effects can be neglected and the bulk value of $R_\textrm{H}$ can be used.

\subsection{Results}

As a worst case scenario we consider a system, where the magnetic field is perpendicular to the heater and has a magnitude $B=B_\textrm{z}=10$~T. We consider a straight 20~mm wide copper plated heater strip with a 40~mm long stainless steel heating station. For the Hall coefficients Alderson \emph{et al.} report $R_\textrm{H,Cu} = -8 \cdot 10^{-11}~\textrm{m}^3/\textrm{C}$ for non-annealed and $R_\textrm{H,Cu} = -6.5 \cdot 10^{-11}~\textrm{m}^3/\textrm{C}$ for annealed copper \cite{PhysRev.174.729}. We take the more conservative value of the non-annealed copper. For stainless steel we use the value $R_\textrm{H,SS} = 3 \cdot 10^{-10}~\textrm{m}^3/\textrm{C}$ measured for Fe-Cr with 20\% chromium \cite{PhysRev.152.498,Hurd}. With these values and 100~A current in a 5~\textmu m copper layer, the Hall voltage across the strip is 16~mV.

\begin{figure}[!t]
\centering
\includegraphics[width=0.9\columnwidth]{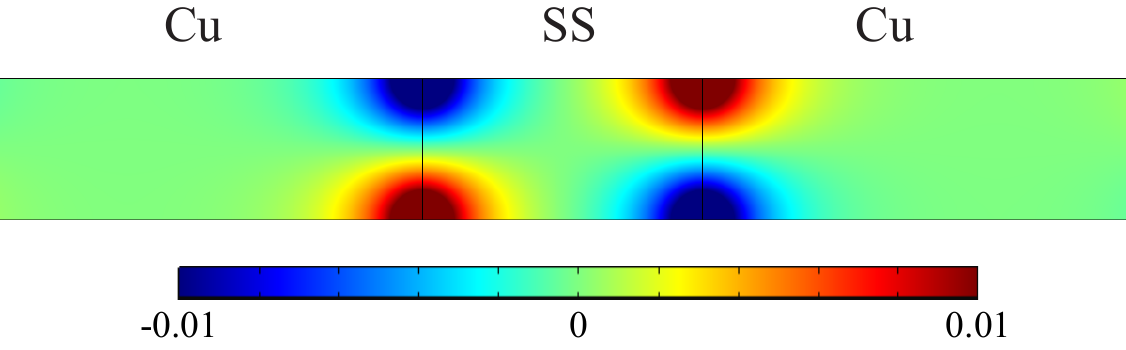}
\caption{Heating density variation due to the Hall effect in a copper plated quench heater strip with a stainless steel heating station in the middle. This is the result of a 2D simulation with a perpendicular magnetic field $B_\textrm{z}=10$~T. The range of $\delta$ is limited to $\pm 1$\%. The maximum variation is $\pm 7.5$\%.}
\label{Fig_HallEffect}
\end{figure}

Fig.~\ref{Fig_HallEffect} shows the results of a 2D simulation for the variation in the current density squared, $\delta (x,y)$, of Eq.~(\ref{Eq_RelativeJSquare}).
In this plot the range of $\delta$ is limited to $\pm 1$\% to show the shape of the heating power density distribution more clearly. The maximum variation in the power density is $\pm7.5$\%. We see that in the bulk, far away from interfaces, current is uniformly distributed. There is a small voltage difference between the upper and the lower sides corresponding to the Hall voltage, but the current is uniform. In the steady state situation the Hall field exactly cancels the magnetic forces and current flows along the same path as it would without the magnetic field. But where two materials with different Hall coefficients meet, there are noticeable differences in the current density, which can potentially cause trouble. This is clearly seen in Fig.~\ref{Fig_HallEffect}. Intuitively these current density patterns can be understood so that the Hall voltages in the two materials are different, which tends to create a voltage difference across the boundary. This voltage drives current loops, which amplify the total current density on one side and decrease it on the other.

\begin{figure}[!t]
\centering
\includegraphics[width=\columnwidth]{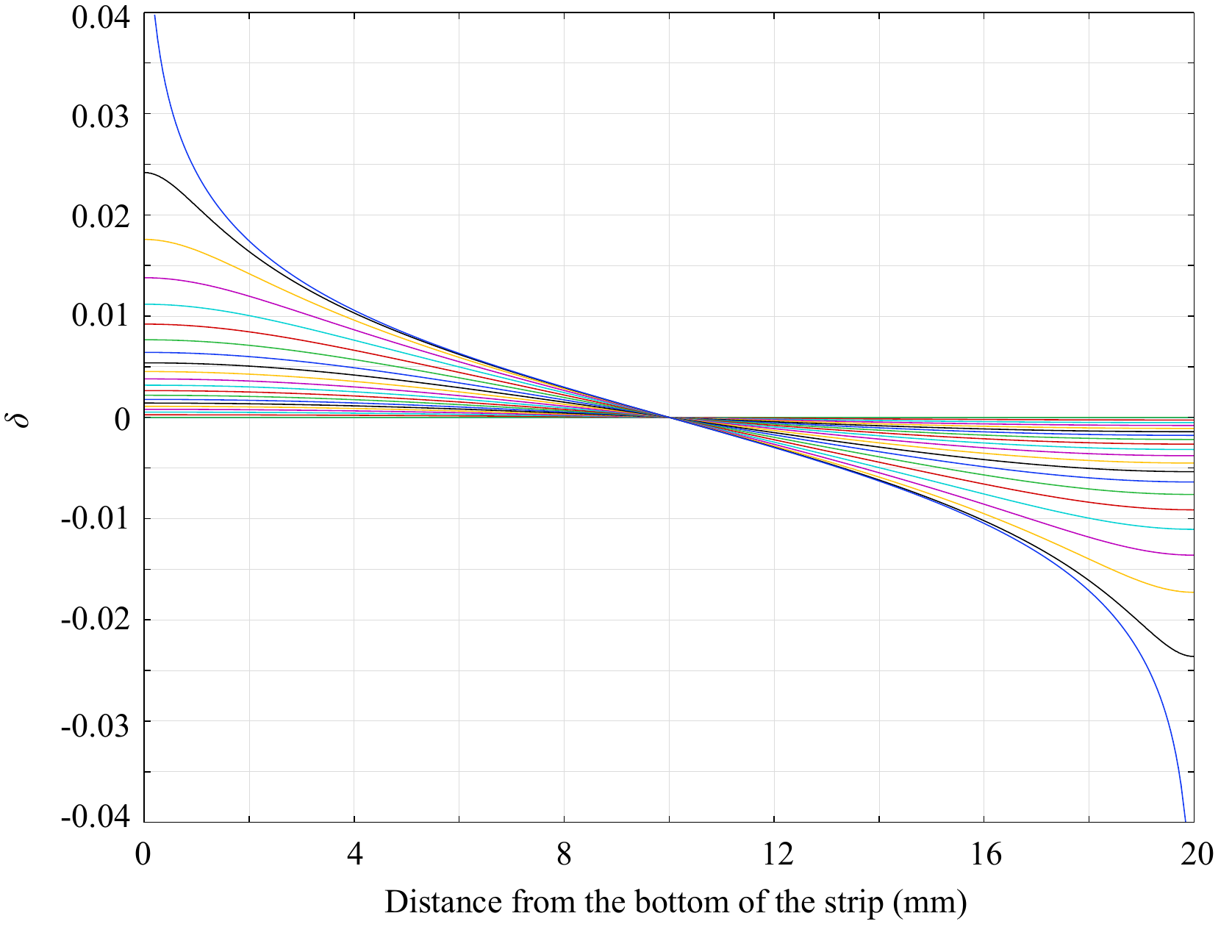}
\caption{Relative heating power density due to the Hall effect in stainless steel along vertical cut lines of different distances from the copper-steel boundary. These are the results of 2D simulation with a perpendicular magnetic field $B=B_\textrm{z}=10$~T. At the boundary the variation reaches $\pm 7.5$\%.}
\label{Fig_HallEffectPlot}
\end{figure}

In Fig.~\ref{Fig_HallEffectPlot} we look at the variation in current density squared in the stainless steel in more detail, plotting $\delta$ along vertical lines with different distances from the steel-copper boundary.
It is seen that fairly quickly the variation in current density is diminished as we move away from the interface. The maximum difference in power density is 7.5\% from the nominal, but occurs only at the very boundary of the two materials.

Since copper-plated quench heaters are really three-dimensional objects with thin layers of metals on top of each other, one may question the applicability of the 2D approximation, even though intuitively the 2D simulations should capture the essential details. To be sure that this approximation does not overlook some important features, we also simulated the entire 3D geometry of a copper-plated stainless steel strip shown in the upper part of Fig.~\ref{Fig_QHGeometry}. Fortunately the results were as expected, almost identical to those of the 2D simulations. Only very close to the boundaries there are tiny deviations. The current distribution becomes essentially the same as in 2D within micrometers from the interface. The minuscule difference is not even produced mainly by the Hall effect, but instead by the fact that the current must shift from the upper layer of copper to the lower layer of stainless steel.

In addition to the possible variations in the heating station current density, the actual 3D geometry brings about another difference to the 2D case. Because of the difference in the Hall voltages between stainless steel and copper, there is also a current loop flowing around in the copper and steel, which is transverse to the main current in the heater. We can easily estimate the magnitude of this current analytically. Most of the main current flows in the copper and its Hall voltage dominates. The ratio of the Hall fields is $E_\textrm{H,Cu}/E_\textrm{H,SS}=(R_\textrm{H,Cu}/R_\textrm{H,SS})(\rho_\textrm{SS}/\rho_\textrm{Cu})\approx 230$. We can therefore consider the Hall voltage of copper being simply short-circuited across the underlying the stainless steel layer. We find the ratio of the transversial current in steel and the main current in copper to be $J_\textrm{SS,x}/J_\textrm{Cu,y}=R_\textrm{H,Cu} B_\textrm{z}/\rho_\textrm{SS}$. For $B_\textrm{z}=10$~T, this ratio is about $J_\textrm{SS,x}/J_\textrm{Cu,y}\approx 0.001$ and thus the transversial current can be neglected in heater performance considerations. The same value for the ratio was seen in the numerical simulations. One should note, however, that for other metal combinations this result can be quite different and the transversial current loops could be important. They would cause heating throughout the quench heater strips, also in regions where heating is not desired, and thus reduce the efficiency of the heating stations.

Above we have considered a magnetic field, which is perpendicular to the heater strip. It is also possible, and for a $\cos \theta$-block magnet coil configuration perhaps even more likely, that the field is not perpendicular to the heater strip, but along it (but still perpendicular to $J$ to produce a Hall voltage). In this case the dimensions of the heater in the relevant direction are much smaller than in the perpendicular case and there no longer exist interfaces of materials where the Hall voltages have a bad mismatch. Therefore, the Hall effect is insignificant. We confirmed this by performing a 2D simulation in the $x$-$z$ plane of Fig.~\ref{Fig_QHGeometry}.

Finally we note a small computational observation. In the 3D simulations it was noticed that the iterative solver had difficulties with convergence. Therefore a direct solver was used, but this requires a large amount of memory. In the 2D simulations the iterative solvers worked without problems.

\section{Conclusion}

Current distributions should be taken into consideration when designing quench heaters, especially if the geometry of the heater is complicated. With copper plating it is possible to correct for some non-uniformities in the current density. Variations in electrical conductivity due to varying magnetic field are not large enough to cause problems. The Hall effect can clearly be more significant, but in our case still manageable with less than 8\% differences in heating from the average. Such differences exist only very close to the boundary of the two materials and rather quickly become uniform in the heating station.

In the future, magnets with higher fields, beyond 20~T, are expected to be developed \cite{Future20TDipole}. If quench heaters are used to protect such magnets, the influence of the Hall effect should be re-evaluated. Higher fields, of course, have the potential for a stronger Hall effect. The details of the analysis depend on the field orientation and the geometry and materials of the quench heaters.

\section*{Acknowledgments}

The author would like to thank E.~Todesco, M.~Marchevsky, P.~Ferracin, and L.~Rossi for useful discussions.


 
\bibliographystyle{IEEEtran}
 
\bibliography{IEEEabrv,References}

\vfill

\end{document}